\documentclass[preprint,epsfig]{revtex4} 
\pdfoutput=1
\usepackage{graphicx} 
\usepackage{subfigure} 
\usepackage{amssymb, amsmath}  
\DeclareMathSizes{7}{11}{3}{3}  
\usepackage{textpos}  
\usepackage{hyperref}
\hypersetup{colorlinks=true, citecolor=blue, urlcolor=blue, linkcolor=blue}

\begin{document}  
\begin{textblock*}{\textwidth}(0cm,0cm)
{{\it This is the accepted author version of the article published in} Physica B vol. 407, pp. 4056-4061 (2012). The version of record is available from the publisher's website from the link: \url{https://doi.org/10.1016/j.physb.2012.01.092}. $\copyright$ 2012 Elsevier. This manuscript version is made available under the \href{https://creativecommons.org/licenses/by-nc-nd/4.0/}{CC-BY-NC-ND 4.0 license}.} 
\end{textblock*}   
\vspace{6cm} 
\title{Photonic crystals as metamaterials} 
\author{S. Foteinopoulou\footnote[1]{Corresponding author: Tel:+44-1392-722101, Fax: +44-1392-264111, e-mail: S.Foteinopoulou@exeter.ac.uk}}
\address{School of Physics, College of Engineering, Mathematics and Physical Sciences (CEMPS), University of Exeter, Exeter, United Kingdom\\}
\begin{abstract}{\vspace{1cm} The visionary work of Veselago had inspired intensive research efforts over the last decade, towards the realization of man-made structures with unprecedented electromagnetic (EM) properties. These structures, known as metamaterials, are typically periodic metallic-based resonant structures demonstrating effective constitutive parameters beyond the possibilities of natural material. For example they can exhibit optical magnetism or simultaneously negative effective permeability and permittivity which implies the existence of a negative refractive index. However, also periodic dielectric and polar material, known as photonic crystals, can exhibit EM capabilities beyond natural materials. This paper reviews the conditions and manifestations of metamaterial capabilities of photonic crystal systems.}
\end{abstract} 
  
\maketitle 
\par
{\bf 1. Introduction}
\vspace{0.2cm}
\par
Veselago's visionary proposal \cite{veselago} in 1967 entailed the fictitious at that time possibility of materials with a simultaneously negative permittivity and permeability. He showed that these materials would demonstrate negative refraction and possess negative refractive index, and also support EM propagation with antiparallel energy and phase velocities (backward waves \cite{tretyakov}). The two pioneering works of Sir John Pendry and co-workers \cite{pend1, pend2} in the late 90's opened up the possibilities for tailoring the effective plasma frequency of a composite wire medium \cite{pend1} and artificial magnetism in metallic resonator structures and have set the stage for the realization of the first negative refractive index metamaterial at GHz frequencies by David R. Smith and co-workers \cite{drs} in 2000. The emerged field of photonic metamaterials now encompasses a variety of theoretically inspired and experimentally realized metallic nanostructures that are functional all the way up to visible frequencies \cite{souk_review, shal_review, shal_book}. Such artificial meta-structures \cite{souk_review, shal_review} possess exotic effective constitutive parameters, i.e. permittivity, $\varepsilon$, and permeability $\mu$ in some or all propagation directions, which are not available in natural  materials. These typically involve a negative effective permeability in frequencies ranging from THz to visible spectrum, which when combined also with a negative effective permittivity lead to an effective negative refracted index and left-handed (backward) EM propagation. Later, metamaterials with effective chiral constitutive relations were also conceived, demonstrating capacity for strong optical activity \cite{chir_zhel, chir_souk, chir_weg, chir_zhang}.
\par 
Photonic crystals (PCs) on the other hand, are also exhibiting extra-ordinary electromagnetic responses, uncharacteristic of natural materials. Photonic crystals\cite{joann} are media composed of dielectric, polar or metallic building blocks arranged periodically in one-, two- or three-dimensional, but do not include resonator elements that are typical to magnetic metamaterials \cite{pend2, souk_review, shal_review}. Their striking EM behavior encompasses unusual ultra-refraction \cite{gralak, fel1} known as the superprism phenomenon \cite{kosaka}, negative refraction \cite{gralak, notomi, sf_prl, sf_longprb, sf_nature, sf_prl2, luo}, multifringent effects \cite{kosaka2, russel1, sf_longprb}, collimation \cite{sf_nature, sf_prl2, chigrin, coljoann}, as well as channeling of the dark field \cite{belov_canal}. Without a doubt, these extra-ordinary EM responses of photonic crystals opened up new avenues for beam manipulation\cite{chigrin,trull,krauss} and sub-wavelength control\cite{luo, luo2, sf_prl2, sridhar_nature, belov_canal}. More recently such functionalities have been demonstrated even in one-dimensional\cite{digennaro}  and quasi-crystal\cite{foc_quasicry} arrangements. 
\par
However, despite their curious electromagnetic response, photonic crystals may not necessarily act as metamaterials, in the sense of possessing effective photonic properties. In this article, the behavior of photonic crystals as metamaterials is reviewed. Two different classes of behaviors are identified. In Sec. 2 the discussion focuses on the first class of photonic crystal metamaterial behavior, entailing PCs which may not posses effective constitutive parameters, but their refractive behavior emulates that of a homogeneous medium with a refractive index n. Such photonic crystal systems are functional at free space wavelengths of the order of the structural meta-atom size, typically two to three times larger. Not all PCs are capable to emulate the refractive properties of a homogeneous block; there are stringent conditions enabling such behavior which are discussed briefly in Sec. 2. 
\par
In Sec. 3 the second class of photonic crystal metamaterials is reviewed. These systems do possess effective constitutive parameters. The demonstrated engineered capabilities for these parameters that go beyond natural materials involve: i) Extreme optical anisotropy \cite{busch},-including indefinite permittivity tensors \cite{indef} for frequencies ranging from THz to visible frequencies. An indefinite permittivity tensor leads to the characteristic EM dispersion with a hyperbolic surface of wavenormals. Hyperbolic dispersion is not possible in natural materials and is attracting increasing interest \cite{engheta,engheta2, prbpolar,fang,zhang,noginov, bolta, schill} as it mediates transfer of the dark spatial frequencies of the source, enabling near-field superfocusing \cite{fang, zhang}. ii) Magnetic behavior -including negative permeability-, mainly at THz and mid-IR frequencies \cite{povinell, shvets, bronger, peng, yannop, yannop2, yannop3, felbacq, park}. The aforementioned types of PC metamaterials are functional at free space wavelengths much larger than the structural meta-atom, typically about 10 times larger or more \cite{prbpolar}.
\par
\vspace{0.4cm}
{\bf 2. Photonic crystals as metamaterials with effective propagation properties without effective constitutive parameters}
\par
\vspace{0.2cm}
\par
\begin{figure}[b]
\begin{center} 
\includegraphics[angle=0,width=8.5cm]{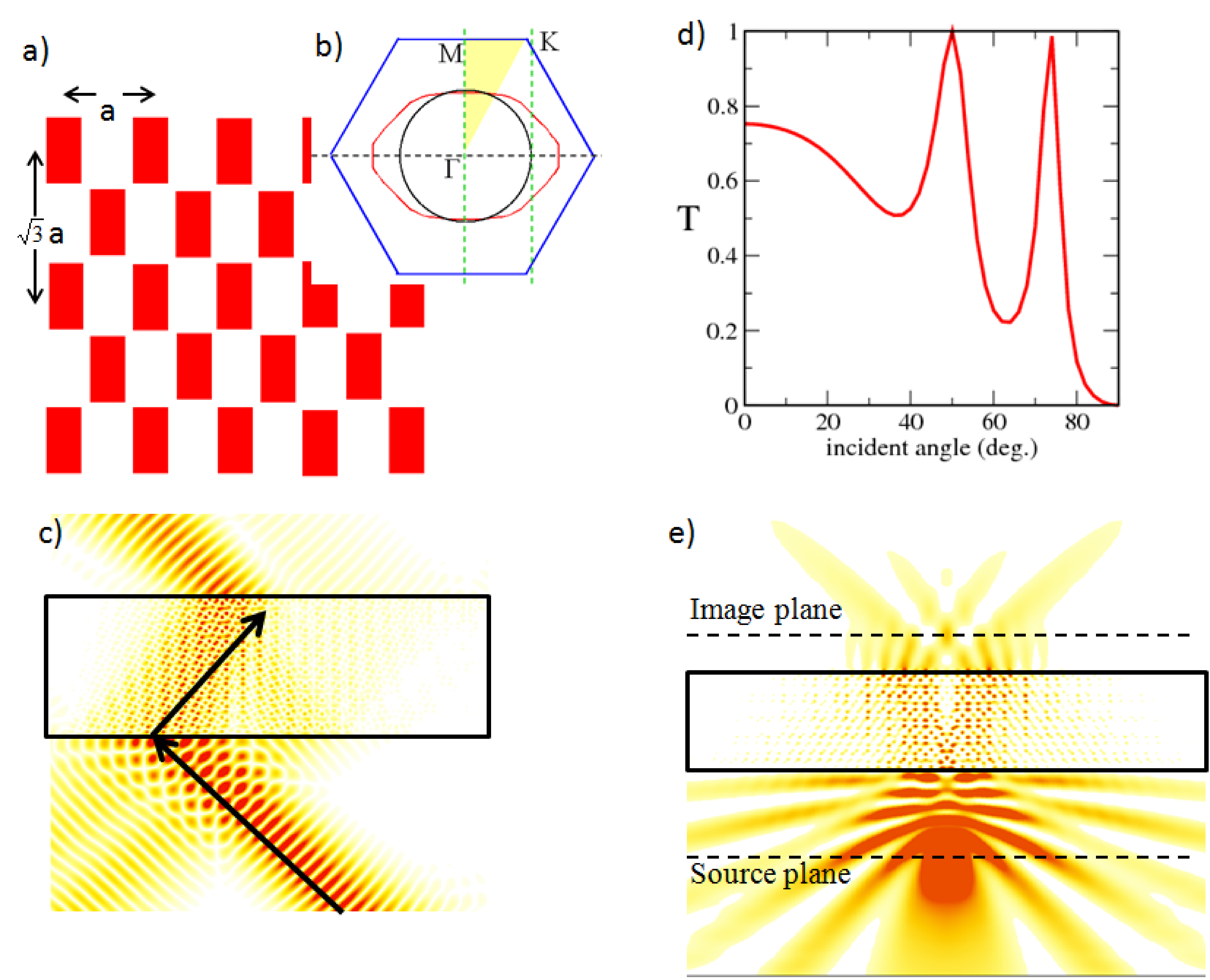}
\caption{(Color online) (a) The PC metamaterial design functional at free space wavelength $\lambda_{free}=2.8985 \hspace{0.25mm}$a. The rectangular rod dimensions are 0.40\hspace{0.25mm}a, and 0.80\hspace{0.1mm}a respectively, with a being the lattice constant. (b) The almost isotropic EFC for the PC metamaterial of $n\sim-1$(compared with the circular EFC in vacuum) (c) Negative refraction at $\sim$ -45 deg. for an incident beam at 45 deg. at the PC metamaterial interface. (d) Transmission versus incident angle for a 12-layer-thick structure. (e) Far field superfocusing through the latter structure, with the incident source placed at $d_s=2.83\hspace{0.25mm} \lambda_{free}$ from the first interface. The image is formed at $d_i \sim 1.33 \hspace{0.25mm} \lambda$. So, $d_s+d_i \sim 12.06$\hspace{0.25mm}a, falling close to the the expected result of 12\hspace{0.25mm}a from Pendry's perfect lens formula.}
\end{center}   
\end{figure}
\par
Photonic crystals generally demonstrate highly complex refractive behavior that is typically multifringent. The native modes of propagation in these systems are the so called Floquet-Bloch waves (FB waves) \cite{brill}, which simultaneously satisfy Maxwell's equation and Bloch's theorem \cite{kmh, sf_longprb}. The FB wave consists of a sum of infinite number of plane waves. However such sum represents a unique propagating mode characterized by a certain energy propagation velocity ${\bf v_e}$; in other words, the FB wave is an entity \cite{sf_longprb}. It is this plane wave sum character of the FB wave that is responsible for the characteristic ``wiggly" phase fronts of EM propagation within a PC medium. Multifringent behavior in PCs arises from the simultaneous coupling of many FB waves under certain conditions of illumination. However, the pioneering work of Notomi in 2000 \cite{notomi} showed that it is possible, in PCs with a high refractive index contrast, to have coupling to a single only FB wave. Furthermore, he showed that at the plane of incidence such PCs possessed a circular contour of wavenormals, just like homogeneous conventional dielectrics do. He termed such contours of wavenormals as Equi-frequency contours (EFC). The refractive index corresponding to such contour was simply, $|n_p|=c k/\omega$, with k being the radius of the EFC contour in wave vector space. This refractive index would yield correctly the magnitude of the refracted angle of a beam hitting the PC interface from Snell's law. Strikingly, refraction was positive for positive slope bands, where ${\bf v_g} \cdot {\bf k}>0$ (with ${\bf v_g}$ being the group velocity and ${\bf k}$ the wave vector and negative for negative slope bands where ${\bf v_g} \cdot {\bf k}<0$ \cite{notomi, voula2}. Note, in dielectric non-dispersive PCs, the group velocity is equal to the energy velocity; ${\bf v_g}$ essentially represents the direction of the Poynting vector, {\bf S} (but averaged within the structural unit cell) \cite{sf_longprb}. It so becomes clear that Notomi's PC structure emulated in its refraction characteristics a left-handed isotropic metamaterial with a negative refractive index. Later, S. Foteinopoulou et al. \cite{sf_longprb} also showed that the group index, $n_g$, yielding the magnitude of the propagation velocity ($v_e=c/|n_g|$) of an EM wave of frequency $\omega$ relates to the phase index, $n_p$ as:
\begin{equation}
n_g=\omega \frac{d |n_p|}{d\omega}+ |n_p|
\end{equation}
just like in a homogeneous material. 
\par
These works implied it is possible for dielectric PCs to emulate, as far as refraction and propagation velocities are concerned, a homogeneous medium with a refractive index $n_p(\omega)$, which can take negative values. In these respects, such PCs act as metamaterials. A plethora of works focused on the quest and analysis of PC designs possessing such capacity demonstrated by almost isotropic surfaces of wave normals (or contours for 2D structures) \cite{sf_prl,foc_quasicry, voula2, luo3, kempa1, kempa2, slhe, prather, efs_opal, rabia}. An example of such PC metamaterial studied by R. Moussa et al. \cite{rabia} is depicted in Fig. 1(a) consisting of alumina rods in air and operating for electric field parallel to the rods axis. The associated EFC is shown in Fig. 1(b) and compared with that of vacuum, for the operational frequency that corresponds to a free space wavelength equal to 2.8985 a,- with a being the lattice constant designated in Fig. 1(a). Fig. 1(c) demonstrates the expected negative refraction at $\sim -45$ deg. for an incident beam of 45 deg., since the refractive index $n_p$ is close to -1.
\par
However, although isotropic EFSs (or EFCs in 2D structures) manifest the possibility for a metamaterial behavior (with respect to refraction and propagation), there are several hovering caveats one must take careful consideration of. The extensive analysis of S. Foteinopoulou et al. \cite{sf_longprb} in various 2D PCs revealed that such metamaterial behavior is subject to the satisfaction of several stringent conditions. These are determined by requiring single beam propagation emanating from direct refraction and not umklapp coupling. Meeting all these conditions entails careful engineering, which in general restricts metamaterial behavior only for certain interface cuts which must be along symmetry directions. Band regions were several bands are present or one band that varies non-monotonically with wavevector lead to multifringence and are so inappropriate for PC metamaterial designs. 
\par
Having the metamaterial capacity of effective refraction and propagation makes these PCs highly attractive for a multitude of applications that rely on such properties. The advantage is that unlike their homogenized resonant metamaterial counterparts, PCs are inherently lossless; so there is no loss of EM energy during propagation. The disadvantage is that these are not a true homogenized medium, thus even structures with $n_p= -1$, can have large reflections \cite{he_bloch}. A high number of works with PCs falling in this category have focused on emulating Pendry's perfect lens \cite{pendrylens} as far as the propagating components of the input source are concerned \cite{foc_quasicry, kempa1, kempa2,  prather,  rabia}. An example is shown in Fig. 1(e) for the design of the work of Ref. \cite{rabia} seen in Fig. 1(a), which was conceived to satisfy the stringent single beam metamaterial conditions at lower frequencies were reflections can be considerably smaller. Indeed, as shown in Fig. 1(d) a high transmission is found for almost all incident angles up to 70 deg. through a structure of 12 rods along the propagation direction. Note the superfocusing demonstrated in Fig. 1(e) is in the far field with the source placed at a distance 2.83 times the free space wavelength from the first interface.
\par
The potential for exploitation of effective propagation properties is enormous and goes beyond superlensing applications. Recently, it has been proposed that PCs  can have a zero phase index\cite{chan_dirac, panoiu} emulating zero refractive index materials \cite{engheta2}. Also, in an arrangement of slow varying building block size they lead to a spatial varying phase index which can be exploited for superbending \cite{cassagne1}, photonic mirage effects \cite{cassagne2, cassagne3} and transformation optics devices such as beam shifters or cloak-like scattering \cite{urzhumov}. Accordingly, in view of their inherently lossless properties such photonic crystal metamaterials are highly attractive for these type of applications. The considerable challenge to be met is optimization of in-coupling efficiency. Surface texturing seems a promising avenue in this direction \cite{rabia, slhe_ter}.
\par
\vspace{0.4cm} 
{\bf 3. Photonic crystals as metamaterials possessing extra-ordinary effective constitutive parameters}
\par
\vspace{0.2cm} 
\par
\begin{figure}[t]
\begin{center} 
\includegraphics[angle=0,width=5.5cm]{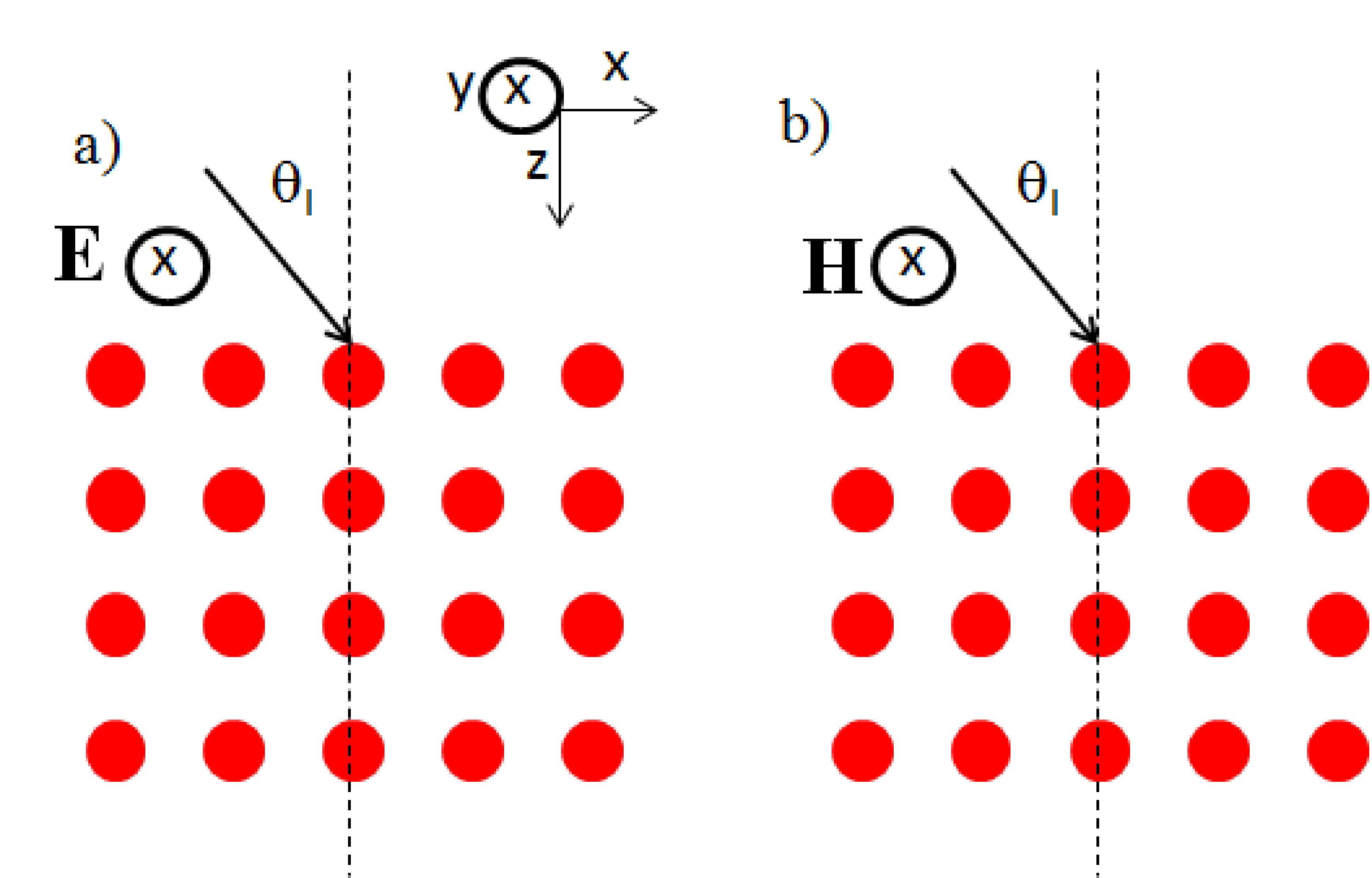}
\caption{(Color online) 2D photonic crystals with infinitely long rods along y. In (a) E-polarization incidence is shown. In (b) H-polarization incidence is shown}
\end{center}   
\end{figure}
This class involves photonic crystals in the deep sub-wavelength regime, where they act as a homogenized bulk medium. Even conventional dielectric 2D PCs (translational symmetry along the y-direction) can lead to extra-ordinary optical anisotropy by far beyond the possibilities in natural media. This is because typically 2D PCs follow a field averaging effective medium for illumination with electric field along y [E-polarization, see Fig. 2(a)], and Maxwell Garnett \cite{maxgar} effective medium theory for magnetic field along y \cite{busch} [H-polarization, see Fig. 2(b)]. Then the whole PC structure acts like a bulk uniaxial block, having permittivity $\varepsilon^E$ in the directions along the optical axis (y), and permittivity $\varepsilon^H$ perpendicular to the optical axis, where $\varepsilon^E$ and $\varepsilon^H$ represent the field-averaging and the Maxwell-Garnett values respectively. Indicatively, the optical anisotropy $|\varepsilon^H-\varepsilon^E|/\varepsilon^E$ for a square PC lattice of Si rods in air and filling ratio of $f=0.30\%$ would be about $60\%$. Evidently, 2D PC composites in the effective medium regime provide a platform for extreme engineering of optical anisotropy. 
\par
How extreme? Lately, it was demonstrated that when plasmonic rods are used as the PC building blocks which have a negative permittivity at optical frequencies, the anisotropy can be so extreme that permittivity is negative along the optical axis and positive perpendicular to that \cite{engheta, engheta2, fang, zhang}. Very recently, S. Foteinopoulou et al. \cite{prbpolar}, explored the possibility to transfer such possibility in the THz and mid-IR regions, where dire need for optical set-up components exists, with the incorporation of polar instead of plasmonic materials. All the aforementioned structures possess an unusual hyperbolic surface of wave normals for the extraordinary mode, which is not encountered in natural media. The response of such PC metamaterials to the ordinary mode is metallic-like with typically a small skin depth, rendering the extra-ordinary mode with hyperbolic dispersion as the dominant mode\cite{prbpolar}.
\par
Not all PC composites that behave as effective metamaterials are subject to the field-averaging/Maxwell Garnett relations, which generally apply for extremely subwavelength meta-atoms. More complicated effective medium theories have been developed by various groups \cite{yannop4, sebas, nariman, halevi, kyria} to extend the frequency range of their validity, including theories predicting magnetic behavior from non-magnetic inclusions \cite{yannop}. Magnetic behavior in such composites emanates from the Mie resonances on the individual meta-atom building blocks \cite{yannop, park}. When the refractive index contrast between the structural building blocks and matrix is very high then such magnetic behavior can be so strong that it enables a negative effective permeability $\mu$ \cite{yannop, yannop2, yannop3}.   In 2D structures effective negative permeability has been reported for a certain direction for high index dielectric \cite{felbacq} and polar material composites \cite{povinell, shvets, bronger, peng, felbacq, park}.
\par
As an example of the possibility of magnetic behavior the effective constitutive parameters for a 2D square PC array of LiF circular rods, -3 $\mu$m in diameter, in a NaCl matrix and spaced 5 $\mu$m apart are shown in Fig. 3. We observe versus free space wavelength, the effective refractive index n [in (a)], the effective permittivity $\varepsilon$ [in (b)] and the effective permeability $\mu$ in [in (c)] for E-polarized waves and normal incidence [$\theta_I=0$ in Fig. 2(a)]. The solid line represent the real part and the dotted lines the imaginary part of the aforementioned quantities. In other words, the effective permittivity represents the $\varepsilon_y$ element of the permittivity tensors and the effective permeability represents the $\mu_z$ element of the permeability tensor, where the direction $x$,$y$ and $z$ are defined in Fig. 2. A very strong, yet non-negative magnetic behavior is seen around 40 $\mu$m, where there is a high permittivity contrast between rods and background of about 50 to 1. Fig. 3(d) shows the real part of the permittivity as a function of free space wavelength for the LiF rods (dotted line) and the NaCl matrix (solid line). Notice, that clearly the strength of the magnetic behavior increases with the permittivity contrast between rods and matrix.
\par
\begin{figure}[t]
\begin{center} 
\includegraphics[angle=0,width=8.5cm]{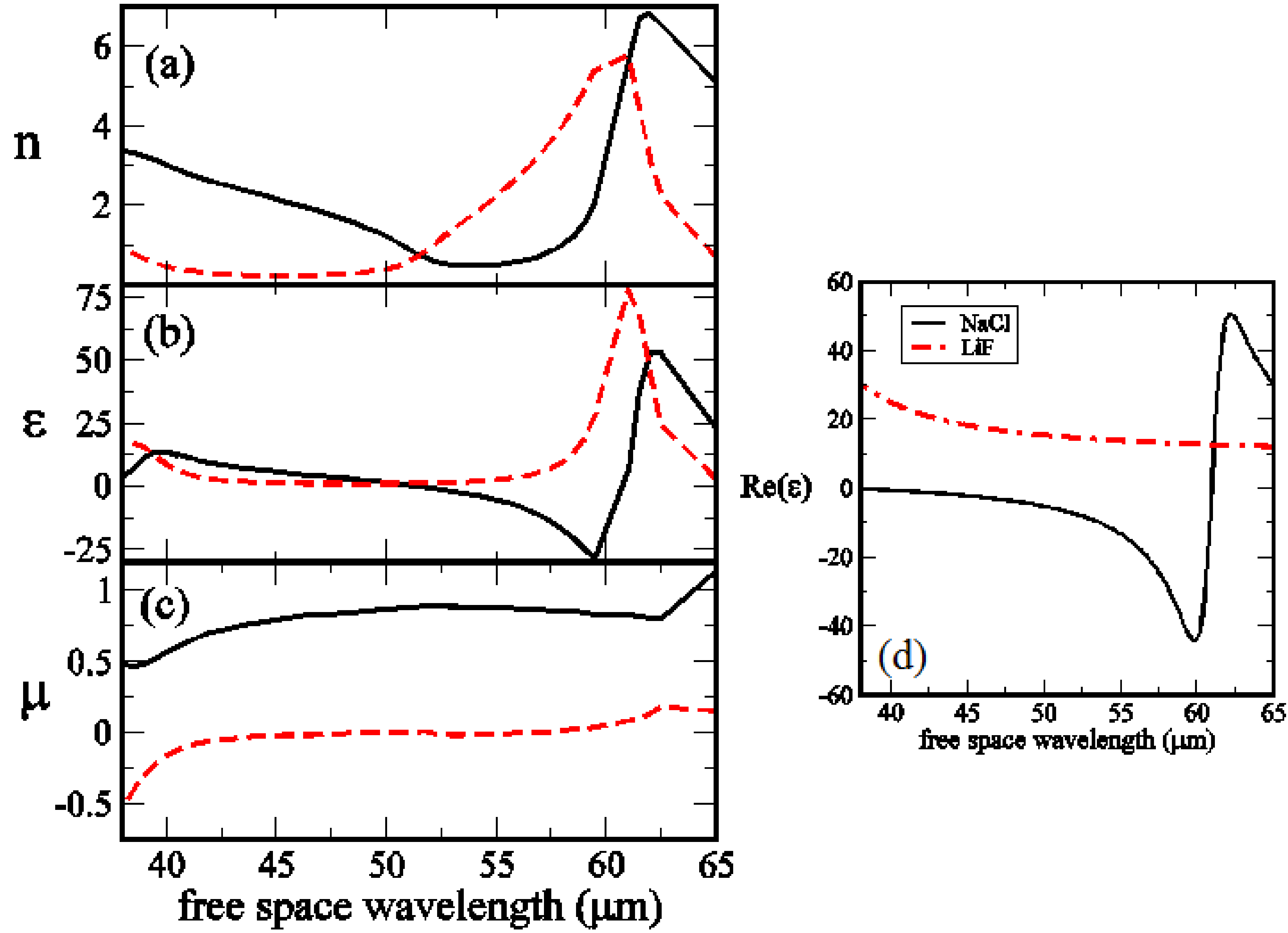}
\caption{(Color online) Retrieved refractive index [panel (a)], permittivity function [panel (b)] and permeability function [panel (c)] for E-waves through the LiF/NaCl composite.  The real (imaginary) part of the retrieved parameters is shown as solid (dashed) lines. In (d) the real part of the bulk permittivity for NaCl, and LiF are depicted (solid and dotted lines respectively).}
\end{center}   
\end{figure}
\par
The effective constitutive parameters are determined from an alternative retrieval method that relies on the information contained in $r/t$ ratio, with $r$ and $t$ being the complex reflectivity and transmissivity. In particular it can be shown \cite{prbpolar} that the effective refractive index can be retrieved from:\\
\begin{equation}
n= \frac{c}{\omega L} \hspace{0.2mm} \left({\cos^{-1}{\displaystyle \frac{{\left({\displaystyle \frac{r}{t}}\right)}_{2L}}{{2 \left({\displaystyle \frac{r}{t}}\right)}_L}}} + \hspace{0.2cm} l \pi \right), 
\end{equation}
where $L$ is the thickness of the structure, $\omega$ the frequency of the EM wave, and $c$ the velocity of light and $l$ signifies the branch solution.  The correct branch is identified after performing the retrieval for several thicknesses, $L$ (and shown in panel (a) of Fig. 3). After the correct branch for $n$ is determined, the impedance, $z$, can be obtained from:\\
\begin{equation}
z-\frac{1}{z} = \frac{\left(\frac{r}{t}\right)_L}{\frac{i}{2}\hspace{0.2mm}  \hspace{0.2mm} sin\hspace{0.2mm} ( {n \hspace{0.2mm} \frac{ \omega L}{c}})},
\end{equation} 
by choosing the root that satisfies passivity, i.e. $Re(z)>$0. Then, the effective permittivity and permeabilities depicted in panels (b) and (c) of Fig. 3 respectively can be easily obtained from $\varepsilon= n z$ and $\mu= \frac{n}{z}$.
\par
Retrieval is off-course meaningful only when the structure does behave as a bulk effective medium, and may or may not agree with simple or more developed effective medium models. It is essential to have available characterization tools that test the validity of metamaterial description of the 2D composite. Properly constructed functions of $r/t$ \cite{prbpolar}, can be shown that should be length independent for a homogeneous medium. Thus testing the variance of these for a large ensemble of thicknesses provides a quantitative measure of effective medium validity. This is however possible only for inherently lossless structures as losses make quickly transmission zero. However, in the same work \cite{prbpolar} it was shown that properly constructed functions of $r/t$ show distinct angular profile that serves as a signature of effective medium behavior. One of them, entailing a $sin^2$ signature becomes too sensitive to numerical error for high permittivities and/or very high losses. The second of these constructed test functions, is however robust even under these conditions and demands a flat angular profile as evidence of metamaterial behavior. In particular, if $\theta_I$ is the incident angle, the function $E(\theta_I)$ \cite{prbpolar}
  
\par
\begin{equation} 
E(\theta_I)=\frac{c^2}{\omega^2 L^2} (Im[\Delta(\theta_I)] \hspace{0.2mm} Re[\Delta(\theta_I)]/\pi + l \hspace{0.2mm} Im[\Delta(\theta_I)]), 
\end{equation} 
with
 \begin{equation} 
\Delta(\theta_I)= \left( {\cos^{-1}{\displaystyle \frac{{\left({\displaystyle \frac{r}{t}}\right)}_{2L}}{{2 \left({\displaystyle \frac{r}{t}}\right)}_L}}}  \hspace{0.1cm}\right)_{\theta_{I}},
\end{equation}
\par
and $l$ the order of the branch should be flat. It can be shown that when magnetic behavior is present then, $E(\theta_I)=Im(\varepsilon \mu)$. Such flat profile test for two characteristic free space wavelengths of 45 $\mu$m and 60 $\mu$m is shown in the insets of Fig. 4, where the flat profile of $E(\theta_I)$ can be clearly seen. The calculated values for the flat $E$ for other frequencies are shown in Fig. 4 as diamonds. Notice, the excellent agreement of the latter with the expected $Im(\varepsilon \mu)$ value (solid line in the figure), ascertaining the metamaterial behavior of the structure.
\par
\begin{figure}[!htb]  
\begin{center} 
\includegraphics[angle=0,width=7.5cm]{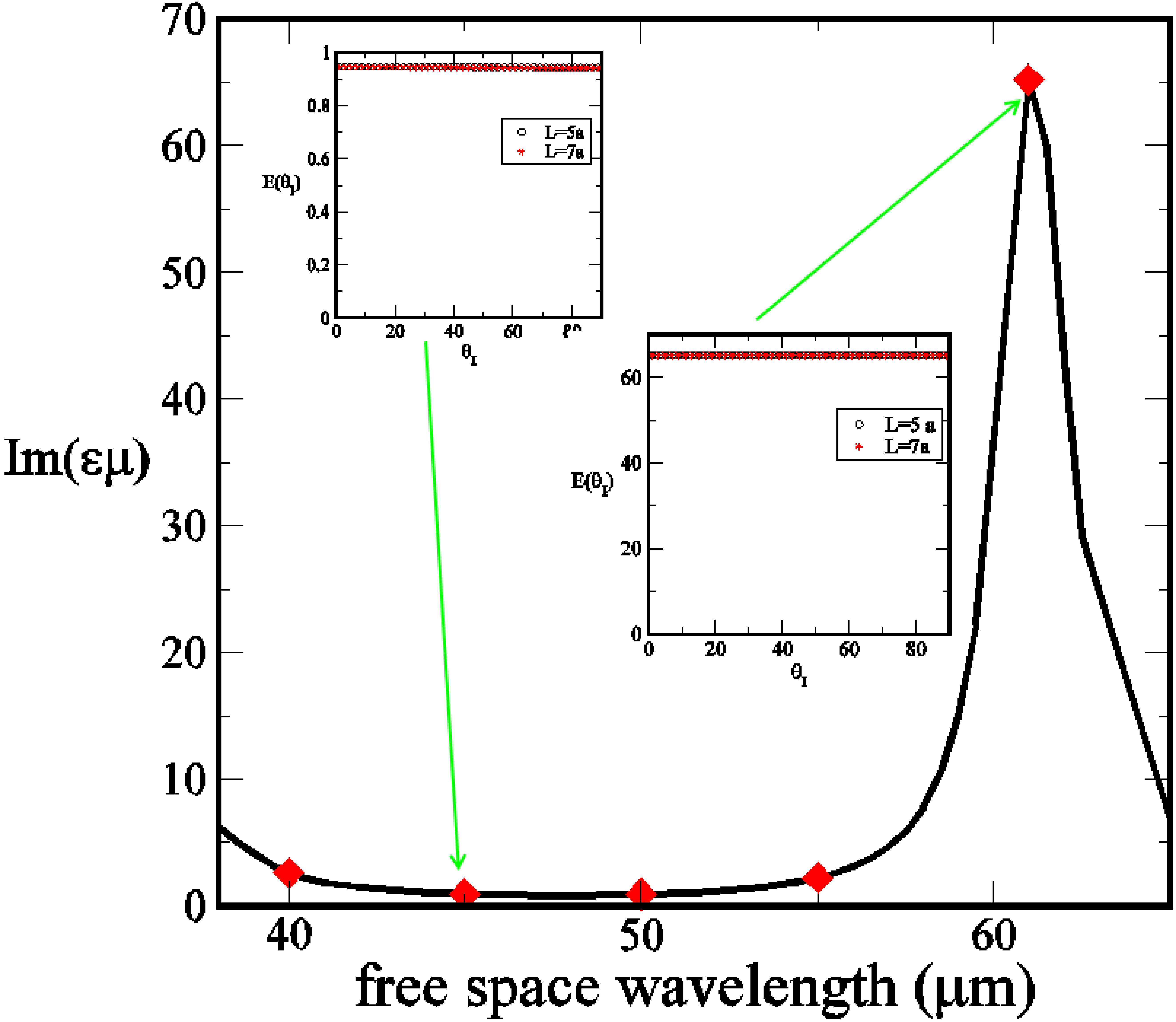}
\caption{(Color online) Flat profile test. The diamonds represent the values of the flat $E(\theta_I)$ function at various free space wavelengths and are compared with the expected value of $Im(\varepsilon \mu)$ (solid line). The actual function $E(\theta_I)$ versus incident angle $\theta_I$ is depicted for two cases in the insets.}
\end{center}   
\end{figure}  
\par
The analysis of Ref. [38] suggests that for 2D composites it is sufficient to check effective medium behavior under the illumination conditions of Fig. 2(a) and Fig. 2(b). If the effective medium criteria apply for these illuminations simulateously, then the entire PC block acts as a uniaxial metamaterial for an arbitrary illumination. 
\par
To recap, extra-ordinary metamaterial behaviors for this class of PCs entails magnetic behavior in some direction(s) and extreme anisotropy with even hyperbolic dispersion. These can be engineered also for the THz and mid-IR regime to open new avenues for beam shaping and manipulation capabilities in these frequencies. Applications exploiting thus far such hyperbolic dispersion are near-field superlenses\cite{fang, zhang} and angle-dependent polarization filtering \cite{prbpolar}. The challenge to be met is constructing composites with a high figure of merit (FOM) given by $Re(k_z)/Im(k_z)$ with $k_z$ being the wave vector along the propagation direction. The results of Ref. [38] are promising in this directions as a FOM exceeding 10 has been reported even in a frequency regime around the polariton resonance of one of the constituents.
\par
\vspace{0.4cm}
{\bf 4. Conclusions}
\vspace{0.2cm}
\par
Metamaterial behavior of photonic crystals was reviewed and two different classes of different characteristics were identified and discussed. In the first class, structures with strict engineering at wavelengths comparable to the structural meta-atom can possess effective propagation properties without possessing effective refractive index. In the second class, deep subwavelength non-magnetic PC composites can demonstrate metamaterial constitutive parameters including magnetic behavior in some or all direction and hyperbolic EM wave dispersion.
\par
\vspace{0.4cm}
{\bf Acknowledgement}
\vspace{0.2cm}

\par
This review paper is written in honor of the sixtieth birthday of Costas M. Soukoulis, celebrated at the recent WavePro Symposium. I would like to take this opportunity to thank him sincerely and heartily for a longstanding, fruitful and inspirational collaboration, which this review aimed to highlight. Part of the work of Ref. \cite{prbpolar} that is reviewed here was done also in collaboration with Maria Kafesaki and Eleftherios N. Economou. 
\par

\end{document}